\PassOptionsToPackage{svgnames,dvipsnames}{xcolor}
\documentclass[journal]{vgtc}

\onlineid{4940}

\vgtccategory{Research}

\vgtcpapertype{Application/Design Study}

\usepackage{paralist}
\usepackage{svg}
\usepackage{amsmath}
\newif\ifhighlight
\highlightfalse

\title{KEditVis: A Visual Analytics System for Knowledge Editing \\ of Large Language Models}

\author{
  Zhenning Chen, Hanbei Zhan, Yanwei Huang, Xin Wu, Dazhen Deng, Di Weng, and Yingcai Wu
}

\authorfooter{
  \item Z. Chen is with Zhejiang University. E-mail: czn@zju.edu.cn.

  \item H. Zhan and Y. Wu are with State Key Lab of CAD\&CG, Zhejiang University. E-mail: \{zhanhanbeiii, ycwu\}@zju.edu.cn.

  \item Y. Huang is with Hong Kong University of Science and Technology. E-mail: yanwei.huang@connect.ust.hk.

  \item X. Wu is with School of Software Engineering, South China University of Technology. E-mail: sexinw@mail.scut.edu.cn.

  \item D. Deng and D. Weng are with School of Software Technology, Zhejiang University. E-mail: \{dengdazhen, dweng\}@zju.edu.cn.
  D. Weng is the corresponding author.

}

\abstract{
  Large Language Models (LLMs) demonstrate exceptional capabilities in factual question answering, yet they sometimes provide incorrect responses. To address this issue, knowledge editing techniques have emerged as effective methods for correcting factual information in LLMs. However, typical knowledge editing workflows struggle with identifying the optimal set of model layers for editing and rely on summary indicators that provide insufficient guidance. This lack of transparency hinders effective comparison and identification of optimal editing strategies. In this paper, we present KEditVis, a novel visual analytics system designed to assist users in gaining a deeper understanding of knowledge editing through interactive visualizations, improving editing outcomes, and discovering valuable insights for the future development of knowledge editing algorithms. With KEditVis, users can select appropriate layers as the editing target, explore the reasons behind ineffective edits, and perform more targeted and effective edits. Our evaluation, including usage scenarios, expert interviews, and a user study, validates the effectiveness and usability of the system.
}


\teaser{
    \centering
    \includegraphics[width=\linewidth]{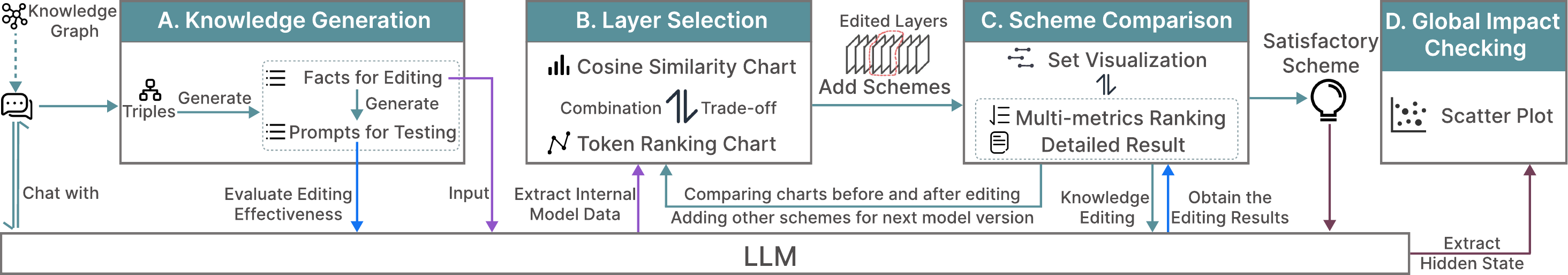}
    \caption{The workflow of KEditVis consists of the following parts: Users start with the LLM chat (they can also start from a knowledge graph, using it to easily generate natural language for the LLM chat) to observe the model's answers to factual questions and determine the editing object. Then: (A) Users generate facts for editing and prompts for testing; (B) Users analyze and select layers as the editing target based on our designed model layer visualizations, including cosine similarity charts and token ranking charts; (C) Users perform comparative analysis of different layer selection schemes, where the layer range relationships between different schemes are presented using set visualizations, and the editing results include both overall outcomes (multi-metrics ranking) and detailed results; (D) After users select a satisfactory editing scheme, they finally use scatter plots to examine the global impact of the edits on the model.
    }
    \label{fig:system-workflow}
}

\graphicspath{{figs/}{figures/}{pictures/}{images/}{./}}

\usepackage{tabu}
\usepackage{booktabs}
\usepackage{lipsum}
\usepackage{mwe}

\usepackage{mathptmx}

\begin{document}
\maketitle

\section{Introduction}

\begin{figure}[tbp]
    \centering
    \includegraphics[width=\columnwidth]{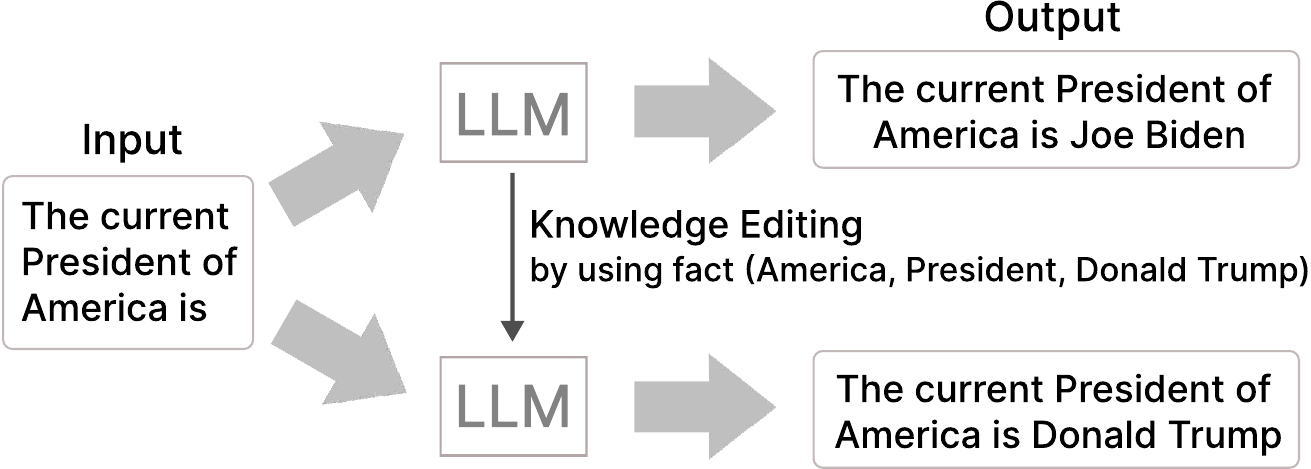}
    \caption{Example of knowledge editing.}
    \label{fig:knowledge-editing}
\end{figure}

Large Language Models (LLMs), pretrained on extensive and massive datasets, demonstrate remarkable capabilities in factual question answering.
However, the accuracy of LLMs' responses is limited by the pretrained datasets, which may contain outdated and inaccurate information, making it necessary to update LLMs from time to time to maintain their knowledge accuracy~\cite{meng2022mass, yin2024history}.
\textit{Knowledge editing} techniques have emerged as an efficient and effective approach for correcting factual information in LLMs, requiring much less computational resources and being less prone to overfitting than traditional fine-tuning, while offering greater accuracy than prompt-based methods~\cite{wang2024knowledge, zhang2024comprehensive}.
These techniques enable users to modify the weights of specific LLM layers based on new factual information (\autoref{fig:knowledge-editing}), ensuring these knowledge updates persist and inform responses to future queries.

Correcting a specific factual error in LLM responses with knowledge editing typically involves a three-phase workflow.
First, relevant facts are gathered and organized into standardized triplets representing the subject, relation, and target answer (e.g., \texttt{(United States, President, Donald Trump)}) to serve as inputs for the editing process.
Next, techniques like MEMIT~\cite{meng2022mass} automatically modify weights across selected consecutive model layers using these facts. Finally, the effectiveness of edits is assessed using quantitative indicators based on a set of testing prompts comprising different partial sentences about the targeted knowledge, like ``\textit{the current leader of America is ...}''.

However, the current workflow heavily depends on manual identification of relevant facts and test prompts, and automated editing techniques may struggle to target the optimal set of layers, which can result in over-editing, producing abnormal responses, or under-editing, where the model continues to respond incorrectly to slightly variant queries.
Moreover, evaluating editing effects across different layer selections requires more than summary indicators alone, as current automated assessment methods can sometimes yield inaccurate assessments, making it necessary to incorporate manual inspection of detailed outcomes.
Overall, the lack of transparency in editing processes hinders the effective comparison and identification of optimal editing strategies.

The limitations observed in the current workflow motivate us to design and develop an interactive approach to knowledge editing of LLMs.
Two major challenges arise in this process:

\textbf{The selection of appropriate model layers as the editing target}. Different types of knowledge have distinct storage patterns across LLM layers, with each type distributed across multiple layers, thus requiring careful selection of the editing layers based on the facts to be edited to achieve more targeted edits. It is challenging to explain and reason about the knowledge editing process and to identify which layers are most relevant to the knowledge being edited due to the complex structure of LLMs. Furthermore, interpreting behavioral changes in models before and after editing, as well as understanding the impact of a single editing operation on the model, continues to pose challenges.

\textbf{The identification of causes of suboptimal editing results}. Current evaluation methods primarily focus on summarizing various metrics in tabular forms, which only present an aggregate overview of results while lacking the granularity needed for detailed analysis. Without thoroughly examining the relationships between edited facts, editing layers, and the resulting model behavior, users lack strategies to refine their edits. Designing effective visualizations is challenging, as they should be capable of supporting multi-criteria comparative analysis of different editing schemes, as well as helping identify issues in suboptimal results and more effective editing schemes.

Through collaboration with four domain experts, we analyzed domain problems and summarized the design requirements for the visual analytics. Based on this, we proposed KEditVis, a novel visual analytics system for knowledge editing in LLMs, which features an interactive edit view specifically designed to address the above challenges.

To address the first challenge, we designed model layer visualizations based on two representative automatic layer selection approaches~\cite{hong2024interpretability, huang-etal-2024-commonsense} to depict high-probability tokens across layers and highlight appropriate editing scopes,
enabling users to interactively weigh candidate layers and select contextually optimal strategies, as well as analyze model changes before and after editing.
For the second challenge, we leveraged set visualization~\cite{alsallakh2016state} to associate different editing schemes with their selected layers and present the overview and detail results of each scheme, helping users conduct in-depth multi-criteria analysis, perform comparisons across different schemes, as well as identify causes of suboptimal edits and better editing schemes.

To evaluate the effectiveness and usability of KEditVis, we presented two usage scenarios that demonstrate its practical value and the insights gained. We also conducted one-on-one interviews with three experts to gather their feedback. Finally, we conducted a user study with 12 participants to further validate the system's usability, which yielded valuable insights regarding both the system and knowledge editing.

The major contributions of this paper are as follows:

\begin{compactitem}
\item We characterize the problem of interactive knowledge editing by soliciting and compiling user requirements through the collaboration with four domain experts.
\item We develop KEditVis, a visual analytics system for exploring knowledge editing insights and performing more targeted and effective edits. Its effectiveness is demonstrated by an evaluation comprising usage scenarios, expert interviews, and a user study.
\end{compactitem}

\section{Related Work}

\subsection{Knowledge Editing Methods}
Based on previous studies~\cite{yao2023editing, wang2024knowledge, wang-etal-2024-easyedit}, we categorize knowledge editing into weight-preservation and weight-modification methods.

Weight-preservation methods can be divided into two types: external memory approaches~\cite{zhong2023mquake, zheng2023can, madaan2022memory, wu2024updating} that leverage external storage and in-context learning~\cite{dong2024survey} to achieve knowledge editing, and parameter expansion approaches~\cite{hartvigsen2024aging, huang2023transformer, dong2022calibrating, wu2025content} that introduce external parameters while keeping the original model parameters unchanged.
However, as the amount of knowledge requiring editing increases, these methods separately show limitations: higher storage capacity requirements; and increased inference burden and system complexity~\cite{li2025swea}.

Weight-modification methods alter model parameters through global optimization or local modification. Global optimization employs specific and constrained strategies like constrained fine-tuning~\cite{chen2020recall, yu2024melo, ni2023forgetting} and intermediate fine-tuning~\cite{cheng2024editing, mitchell2021fast, de2021editing} -- unlike simple fine-tuning which easily leads to overfitting. However, these approaches remain inefficient due to the large number of model parameters and can easily affect knowledge outside the editing scope.
The local modification methods, which this paper primarily focuses on, represent current mainstream knowledge editing approaches. These methods are built upon the finding that feedforward neural networks are key-value memories~\cite{geva2020transformer}, thus following a locate-then-edit strategy. While early studies such as KD~\cite{dai2021knowledge} and ROME~\cite{meng2022locating} focused on editing individual facts, the introduction of MEMIT~\cite{meng2022mass} enabled the simultaneous editing of multiple knowledge instances.
Subsequently, several methods built on and further improved MEMIT have emerged~\cite{jiang2024neuron, huang-etal-2024-commonsense, li2024pmet, fang2025alphaedit}.
Among them, AlphaEdit~\cite{fang2025alphaedit} is the latest, achieving state-of-the-art performance through minor modifications to MEMIT with only a single line of code added. Our proposed visual analytics system offers generalizability and is not constrained by specific editing methods; it supports the aforementioned mainstream approaches such as MEMIT and AlphaEdit.

\subsection{Visual Analytics for Machine Learning}
There has been an increasing amount of studies leveraging visualization to support machine learning, with the aim of facilitating the understanding and improvement of machine learning models~\cite{yuan2021survey, ruan2024violet, xuan2025attributionscanner, deng2024adversaflow}. We categorize these studies into two types: approaches exploring models from an architectural perspective and those exploring models from a data perspective.
For the former, some studies have utilized interactive methods to reveal the structure of deep neural networks~\cite{kahng2017cti, rauber2016visualizing, pezzotti2017deepeyes, choo2018visual} and DNN-based models, such as CNNs~\cite{liu2016towards, bilal2017convolutional, wang2020cnn}, RNNs~\cite{ming2017understanding, strobelt2017lstmvis, li2015visualizing}, and Transformers~\cite{yeh2023attentionviz}. For the latter, some works start from a data perspective~\cite{wang2024visual}, utilizing visualization techniques to explore and improve both the model and the data associated with it~\cite{wang2019deepvid, he2020dynamicsexplorer, li2024evovis, wang2021investigating, cabrera2019fairvis, li2018embeddingvis, chen2020oodanalyzer, ren2016squares}.

By contrast, our research explores knowledge editing from both model architecture and data perspectives. Since previous work lacks interactive methods that directly target knowledge editing processes, our work offers novelty and usefulness to the field of visual analytics.

\subsection{Multi-Criteria Decision Making}
Multi-criteria decision making (MCDM) is a method that helps decision-makers make informed decisions under multiple conflicting attributes or criteria. We categorize MCDM methods into traditional methods and interactive methods.

Traditional methods consist of Analytic Hierarchy Process (AHP), Fuzzy AHP, TOPSIS, ELECTRE, and Grey Theory~\cite{aruldoss2013survey}. For instance, TOPSIS~\cite{opricovic2004compromise} calculates the distance between alternatives and ideal solutions to rank them.
Visual ranking techniques dominate interactive methods, providing decision-makers with intuitive visualizations of multiple metrics to effectively sort and analyze alternatives. These techniques include tabular-based and glyph-based visualizations. For tabular-based visualizations, ValueCharts~\cite{carenini2004valuecharts} stands as a pioneering work; building upon this foundation, LineUp~\cite{gratzl2013lineup} enables the interactive comparison of multi-attribute rankings through intuitive visualizations of weighted attributes and ranking changes. Glyph-based visualizations use various glyphs to visually compare multiple rankings~\cite{behrisch2013visual, pajer2016weightlifter}.
For example, Behrisch et al.~\cite{behrisch2013visual} used radial node-link glyphs to help users identify patterns across different ordering methods.

In this paper, we also leverage visual rankings to compare editing outcomes across different schemes. We employ set visualization~\cite{alsallakh2016state, lex2014upset, alper2011design, meulemans2013kelpfusion} to display the relationships between schemes, with our algorithm reducing visual clutter to better facilitate comparative analysis.

\section{Requirement Analysis}

\subsection{Background and Problem Formulation}
To understand the current knowledge editing workflow and its limitations, we collaborated closely with four domain experts (E1, E2, E3, E4) over the past year. E1 and E2 are NLP experts with a deep understanding of knowledge editing mechanisms; E3 and E4 are interdisciplinary researchers whose work focuses on visualization for machine learning.
We held regular online meetings with these experts, during which we asked them to share their experience with knowledge editing and envision improved solutions that incorporate interactive approaches. Through this collaboration and a review of relevant literature~\cite{yao2023editing, wang2024knowledge, wang-etal-2024-easyedit, fang2025alphaedit, meng2022mass}, we have summarized the application scenarios, experts' method, and workflow as follows.

\textbf{Application scenarios and research scope}. Knowledge editing ``\textit{is a relatively fine-grained model editing method suitable for making precise adjustments}'' (E2). When editing a small number of facts, knowledge editing methods ``\textit{are less likely to cause overfitting}'' (E2). However, for large-scale editing, these methods may not be superior to traditional fine-tuning, and large-scale knowledge editing ``\textit{can introduce toxicity and potentially lead to model collapse}'' (E1, E2). In practical applications, large-scale editing ``\textit{typically employs traditional fine-tuning rather than knowledge editing methods}'' (E1, E2). Meanwhile, editing a small number of facts is widely common in practical scenarios, such as when occasionally correcting a few errors that are discovered after model deployment (E1, E2, E3).

Based on these observations, we decided to focus this study on small-scale editing (1-100 facts per session).
While the current solution may not scale well for large-scale editing tasks, we note that in these uncommon cases, automated methods could be applied first, and when poor edits are detected, they could be routed to our system.

\textbf{Knowledge editing method}.
Experts pointed out that among the categories of the knowledge editing methods, local modification methods, which follow a locate-then-edit strategy, are currently the most mainstream and representative.
Our literature review also supports this claim~\cite{yao2023editing, wang2024knowledge, wang-etal-2024-easyedit, meng2022mass, fang2025alphaedit}.
Among these methods, MEMIT~\cite{meng2022mass} is the most popular one, while AlphaEdit\cite{fang2025alphaedit} is currently the state-of-the-art. Regardless of the specific method details, they typically operate in two key steps: (1) computing target vectors $z_i$ (i represents the i-th fact) for each fact to be edited, and (2) sequentially updating weights of pre-selected MLP layers identified through preliminary experiments. The pre-selected layers must form a continuous range, and weight updates proceed from lower layers toward higher layers, with each layer's weight change calculated based on both the weights of previously updated layers and the $z_i$ values computed for all facts in the first step. For a more detailed introduction, please refer to the appendix A.

A primary limitation of current editing approaches lies in their layer selection methodology, which relies on \textit{static, model-specific} presets (e.g., in MEMIT, layers 4–8 for Llama3-8B and 13–17 for GPT2-XL), thus ignoring the \textit{fact-specific} contextual signals required for precise edits. In particular, repeated edits on fixed layers may induce cumulative parameter drift -- a primary cause of catastrophic model collapse, urging for interactive approaches that can select the range of layers based on knowledge editing goals.

\textbf{Workflow and limitations}.
Currently, the typical knowledge editing workflow is divided into three phases, each with its own limitations.

First, both editing facts and testing prompts need to be properly constructed. The former should be organized into triplet format, while the latter should be divided into multiple testing categories, each containing several prompts. Experts noted this process ``\textit{requires manually constructing prompts based on the edited fact, which is tedious}'' (E4).

Second, the automatic methods called for knowledge editing face limitations as they ``\textit{make it difficult to target the layers most suitable for editing}'' (E1), leading to over or under-editing. The edited model can easily ``\textit{confuse concepts, produce hallucinations, or output repetitive words}'' (E2). The experts consider the selection of editing layers the most important and flexible hyperparameter. Nevertheless, interpreting LLMs' complex structure is not a straightforward task. It is challenging to extract key model data and to visualize it in ways that effectively guide users toward making reasonable layer selection decisions.

Finally, several overall metrics are used to provide a general assessment of the editing results. However, experts highlighted the limitations of automated evaluation methods, noting that they sometimes incorrectly classify outputs that actually contain correct answers as erroneous. This highlights the necessity and value of human involvement in the analysis process. Overall, due to the lack of transparency in the editing process, it is challenging to compare different strategies and identify the factors that contribute to optimal or suboptimal editing outcomes.

In addition, the experts noted that due to the lack of interactive methods specifically designed for knowledge editing, users cannot effectively explore or refine their editing strategies. Existing research has predominantly focused on algorithmic advancements, leaving users unable to identify or explore potential adjustment directions when edits targeting individual facts or specific editing intentions yield suboptimal results. This lack of flexible interaction mechanisms limits the ability to achieve more effective and targeted knowledge editing in LLMs.

\subsection{Analytical Requirements}
Our goal is to help LLM practitioners perform knowledge editing more precisely and effectively, as well as to explore relevant insights that can benefit the development of knowledge editing techniques. During online meetings, we conducted brainstorming sessions and prototype discussions with domain experts. This iterative process enabled us to refine and identify key user requirements for the visualization system. Ultimately, we summarized the following six user requirements.

\textbf{R1: Convenient and effective generation of facts and prompts.} The experts noted that fact and prompt construction is tedious in typical workflows. The system should automatically generate these elements based on user intent. Users need efficient ways to create facts and comprehensive prompts across multiple categories to evaluate edit effectiveness and potential knowledge interference. The experts also suggested including knowledge graphs as references.

\textbf{R2: Selection of appropriate model layers for editing.} In typical workflows, automatically executing knowledge editing methods fails to target the most suitable layers, leading to poor results.
The system should extract and visualize key model data relevant to the intended edit for informed layer selections. Users should be able to analyze and weigh trade-offs, select appropriate layers, as well as understand model changes by comparing visualizations before and after editing.

\textbf{R3: Comparative analysis across editing strategies.} Due to the lack of transparency in the editing process, it becomes difficult to compare editing strategies and understand poor outcomes. The system should support comparative analysis of different editing schemes.
Users should be able to analyze variations between editing strategies and their corresponding results, gaining practical insights into why certain methods are effective while others produce suboptimal outcomes.

\textbf{R4: Comprehensive presentation of editing results.} Evaluation in typical workflows relies on several overview metrics, the results of which are sometimes not particularly accurate. Thus, users need both overview and detailed visualizations of editing results. The overview visualization provides a summary of evaluation results through core assessment metrics, while detailed visualization requires text visualization to display and compare each output in detail.

\textbf{R5: Iterative algorithm execution and reversible model weights.}
The experts believed that in the process of interactive exploration and execution of knowledge editing, it cannot be concluded with just a single edit. Users should be able to perform multiple edits on a model in succession, with each edit building on previous results. If users discover that an editing attempt has produced undesirable outcomes, they can revert the model weights to before the last edit.

\textbf{R6: Analysis of the global impact of editing on the model.}
The experts noted that semantic conflicts in edited facts may potentially lead to model collapse.
To prevent this, users should be able to observe and evaluate the global impact on the model, and check if any negative effects have occurred.
For providing a more understandable approach for users, the experts suggested analyzing from the perspective of data input-output and whether shifts have occurred.

\section{KEditVis}

\begin{figure*}[t]
    \centering
    \includegraphics[width=\textwidth]{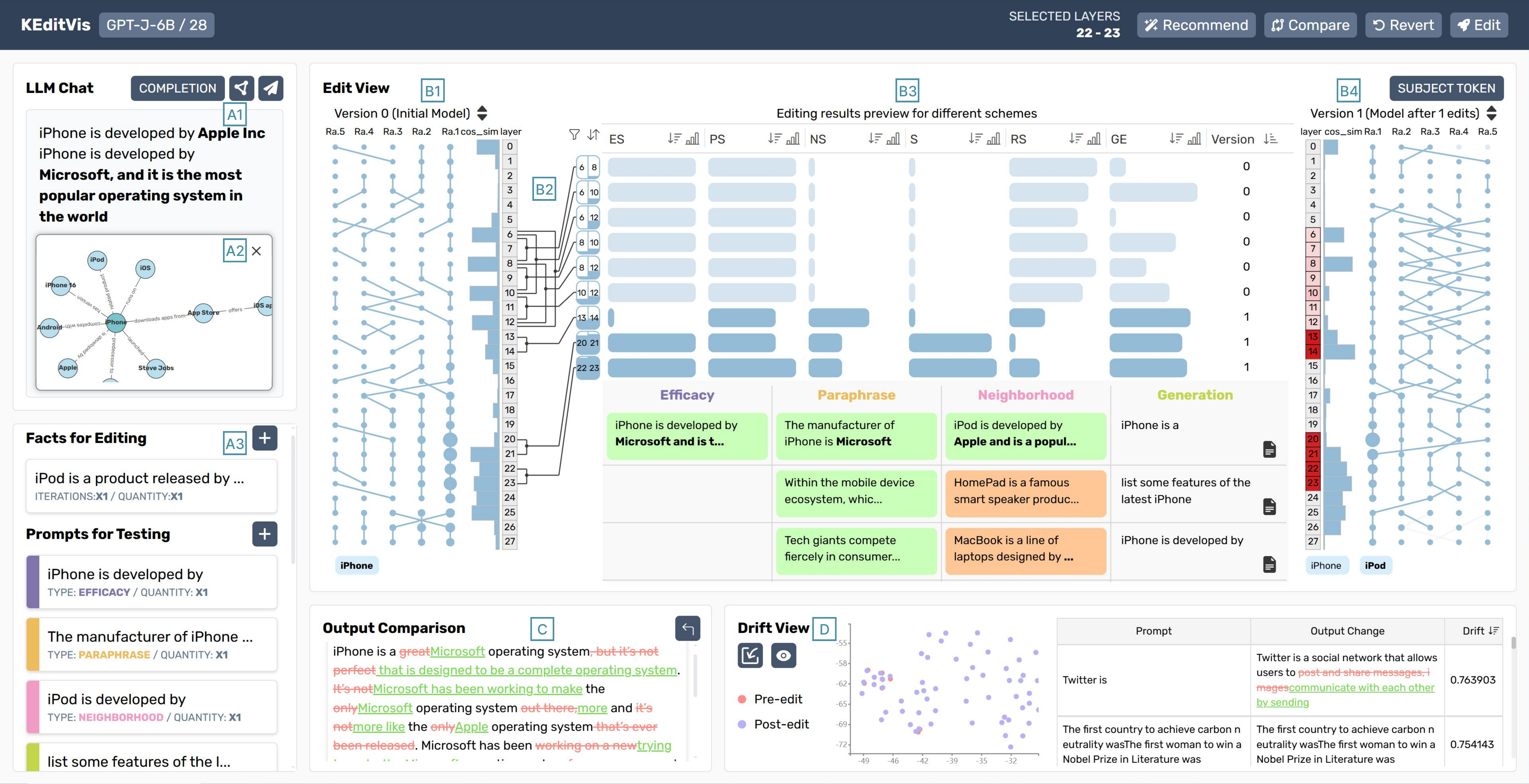}
    \caption{The interface of our visual analytics system KEditVis. The system provides (A1) an LLM chat view for interactive dialogue, (A2) a knowledge graph, and (A3) management of editing facts and testing prompts. The edit view visualizes (B1 and B4) key internal data of the model layers both before and after editing, facilitating the selection of appropriate layers for editing  and comparison of different model versions. The edit view leverages (B2) set visualization to show the relationship between different schemes, and combines (B3) both overview results and detailed results to compare and evaluate the editing outcomes across schemes. (C) The output comparison view employs text visualization to compare the model’s outputs before and after model editing. (D) The drift view visualizes the global impact of edits on the model through a scatter plot.}
    \label{fig:teaser}
\end{figure*}

Based on the requirements, we designed KEditVis, a visual analytics system for helping model experts explore insights and guidance to perform more targeted and effective edits.
The overview of KEditVis's workflow is shown in \autoref{fig:system-workflow}.
The edit view, as the core view of the system, visualizes internal model data to facilitate selecting appropriate layers for editing (\autoref{fig:teaser}B1, B4), providing an understanding of model changes before and after editing (R2).
This view offers multi-dimensional features including presenting different layer selection schemes (\autoref{fig:teaser}B2) (R3), while providing an overview comparison of editing results across schemes and displaying detailed outcomes for comprehensive analysis (\autoref{fig:teaser}B3) (R3, R4).

\textbf{Justification.}
We abandoned designing interpretative visualizations directly for editing algorithms and mechanisms, instead focusing on understanding edits through other perspectives.
This is because LLMs' complexity and algorithmic abstractness make it difficult to explain phenomena solely from algorithms.
Second, such an approach would reduce generalizability and not support different editing methods.

\subsection{Generating Editing Facts and Testing Prompts}

\textbf{Interface Design}. The system provides the LLM chat view (\autoref{fig:teaser}A1) and the knowledge graph (\autoref{fig:teaser}A2). Both the chat view and knowledge graphs can serve as the starting point of a workflow, accommodating two different scenarios. In one case, users arrive with specific questions and directly engage in dialogue with the LLM to examine the answers to factual questions; in the other case, users do not have specific questions in mind and instead use knowledge graphs to search for and investigate potential issues.

The chat view provides two modes: ``Completion'' and ``Rewrite''.
The ``Completion'' mode enables users to examine how the currently selected model (i.e., the model to be edited) completes answers to factual questions, with model-generated content displayed in bold text to distinguish it from user input.
The ``Rewrite'' mode leverages GPT-4 to generate several test prompts (which can be added to the list of test prompts below) based on the user's written sentences.
Moreover, the system manages facts for editing and prompts for testing (\autoref{fig:teaser}A3).

Test prompts are categorized according to different evaluation metrics.
Drawing from prior research~\cite{hong2024interpretability, meng2022mass, huang-etal-2024-commonsense}, the following metrics are employed to assess editing performance:
Efficacy Success (\textbf{ES}) measures whether the model produces edited knowledge when given the original prompt;
Paraphrase Success (\textbf{PS}) tests whether edited knowledge applies when responding to paraphrased prompts;
Neighborhood success (\textbf{NS}) verifies that edits only affect targeted knowledge without changing similar content;
Editing Score (\textbf{S}) is the harmonic mean of ES, PS, and NS;
Reference Score (\textbf{RS}) evaluates alignment between model outputs and reliable references like Wikipedia;
Generation Entropy (\textbf{GE}) assesses output naturalness and language quality after editing.
Higher values indicate better performance for all the above metrics.

As shown in \autoref{fig:teaser}A3, test prompts are categorized with different colored tags according to their evaluation metrics: ``Efficacy'', ``Paraphrase'', ``Neighborhood'', and ``Generation'' (R1). The first three are used to evaluate the ES, PS, and NS metrics respectively, while the last one is used to evaluate both RS and GE.

\textbf{Interaction}.
The knowledge graph, constructed in real-time based on user-input keywords (R1), can generate corresponding natural language questions in the LLM chat view when two nodes are clicked, eliminating manual typing.
Clicking ``+'' above facts or prompts opens a popup window: for facts, users can either add their own or let the system generate multiple synonymous expressions of a given fact (they can be used together in one edit to enhance the model's memory of this knowledge); for prompts, users can add their own or have the system automatically generate various test prompts based on the editing object (R1).
Automated generation eliminates tedious manual work by users.

\subsection{Selecting Layers for Editing}
\textbf{Visual Design}. The edit view integrates two layer selection approaches and displays critical data, including the cosine similarities across different model layers and the five highest probability tokens for each layer, to assist users in making informed layer selection (R2).
These two types of data are visualized through our designed model layer visualizations (\autoref{fig:teaser}B1), including a cosine similarity bar chart and a token probability ranking chart, which are integrated into a unified visualization design as they share the layer axis as a common coordinate.
This visualization design is positioned on both the leftmost and rightmost sides of the edit view (\autoref{fig:teaser}B1, B4), enabling users to compare internal data across different edited versions of the model (i.e., the model after a specific number of edits) (R2).

\textbf{Interaction}.
When double-clicking a fact before editing, it serves as input to the model, and subsequently the corresponding bar chart and ranking chart will be displayed in the edit view.
Users can not only add layer ranges by clicking specific cosine similarity bars or circles in the ranking chart, but also remove previously selected layer ranges by clicking on the scheme (rectangle with numbers).
The ``Recommend'' button automatically suggests potentially effective sub-ranges within the most recently selected layer range (R2). It identifies the 50\% of layers with the lowest cosine similarity values in this range and uses combinations of these as boundaries for new sub-ranges.

\textbf{Layer Selection Approach.}
By surveying recent knowledge editing studies, we identified two approaches from the literature~\cite{hong2024interpretability, huang-etal-2024-commonsense} for dynamically selecting layers for editing. The inclusion of these two approaches helps users balance the advantages of both methods, providing references from complementary perspectives on numerical significance and token-level semantic information. This allows users to make trade-offs based on their specific needs and interests when selecting layers. We refer to these two approaches as the cosine similarity approach and the token projection approach, introduced as follows:

\subsubsection{Cosine Similarity Approach}
When inputting the edited fact into the model, the cosine similarity between the input and output hidden states can be calculated for each MLP layer; layers with lower similarities indicate greater contributions to knowledge processing and are therefore ideal candidates for the starting and ending layers of the editing range~\cite{huang-etal-2024-commonsense}.

\textbf{Design of the cosine similarity bar chart.}
We encode the magnitude of cosine similarity values through the bar length. The mapping between bar length and cosine similarity value follows this function:
$length = L_{\max} \cdot (1 - \tanh(\beta \cdot \text{cos\_sim}))$,
where $L_{\max}$ represents the maximum bar length when the cosine similarity value equals zero. There are two main reasons for choosing this function. First, we need to assign longer bars to represent cosine similarity values closer to zero, which are considered more significant, as longer bars make these important values more readily identifiable to users. Second, we observed that several layers often exhibit cosine similarity values very close to zero, which would result in nearly identical bar lengths if a linear mapping were used, thus making visual discrimination difficult. We set $\beta=6$ to amplify the differences in bar length for these near-zero cosine similarity values, enabling clearer visual distinction.

\subsubsection{Token Projection Approach}
When inputting the edited fact into the model, the five highest probability tokens across each MLP layer can be computed by projecting the representation of the input content's subject token onto the vocabulary space; the range between the two layers where target tokens show the highest probabilities is considered to be processing target knowledge and is typically selected~\cite{hong2024interpretability}.
Additionally, we can use the same method to calculate the highest probability tokens for the input content's last token, which shows the model's probability predictions for the next token output based on the current input.

\textbf{Design of the token probability ranking chart.}
The ranking chart visualizes the five highest-probability tokens at each layer.
We encode token probability values across layers using variable-sized circles, with larger circles indicating higher probabilities, minimal semi-transparent circles representing zero values. Connecting lines link the same token across layers. Hovering over any circle reveals detailed token information while highlighting its complete path and dimming others.

Since ranking charts for the subject token and for the last token can use the same visualization method, the system provides visualization for both, allowing users to switch between them.
The former directly helps select appropriate editing layers, while the latter reveals the model's reasoning path and how it generates the final output tokens step by step.

\textbf{Justification.} We chose ranking charts after evaluating alternatives.
Matrix visualization seemed suitable for discrete token probability data using color intensity encoding, but testing with real data revealed poor representation of token relationships and cross-layer changes (see the appendix B.3.1).
Unlike fixed-position matrices, ranking charts effectively display probability values and track changes across layers while revealing data stability and model information processing characteristics through flexible positioning, better aligning with our requirements.

\subsection{Comparing Different Schemes}

The wireframes (links associating ranges of layers to rows in the table as in \autoref{fig:teaser}B2) and table (\autoref{fig:teaser}B3) in the edit view are designed to evaluate and compare different layer selection schemes (R3).
The wireframe provides the visualization of different layer selection sets, allowing users to explore relationships between different selections, such as intersection and difference relations.
The table features a multi-metric ranking visualization, where longer bars represent higher values.

After clicking the ``Compare'' button, the system generates a preview of knowledge editing evaluation results in the table for different layer selection schemes based on all the test prompts (R3).
In the column header, clicking the sort icon arranges metrics in descending order; clicking the distribution graph icon displays layer-wise metric distributions in the layer column, using red intensity to encode values.

\subsubsection{Layer Selection Visualization with Wireframes}
Since the layer selection scheme includes continuous layers, we visualize the set relationships between schemes using wireframes that connect the first and last layers of each scheme's layer range (R3).
As shown in \autoref{Quantitative-results}A,
when schemes share boundary layers, we divide layer rectangle edges (in the ``layer'' column) into equal segments for different wireframe connections, preventing visual clutter from overlapping wireframes.
The horizontal lengths of different wireframes may vary, further preventing visual clutter from overlapping wireframes.

Displaying wireframes to minimize visual clutter is an NP-hard problem. To generate and update these wireframes in real time, we designed a heuristic visualization algorithm based on a greedy strategy that determines the layout of each wireframe.
First, we generate an ordering for schemes based on their layer scope. Second, we traverse the ordering and calculate each wireframe's horizontal length (ensuring the smallest possible horizontal length while maintaining no overlaps) and the number of division points for each layer rectangle's edge. Third, we traverse each layer and connect each wireframe to its optimal division points to minimize wireframe crossings. This algorithm optimizes wireframes to minimize crossings and overlaps while using the smallest possible number of distinct horizontal lengths. Its worst-case bound of time complexity is only $O(L \times m \log m)$, where $L$ represents the maximum number of layers, and $m$ represents the number of schemes.
The detailed algorithms can be found in the appendix B.2.

\textbf{Interaction}.
As shown in \autoref{Quantitative-results}A, clicking the sort icon sorts schemes based on the intersection points of wireframes and dotted line $x_1$, thereby eliminating crossing lines between the two dotted lines $x_1$ and $x_2$ (see the crossing lines in \autoref{fig:scenario1}C before sorting).
Furthermore, to help discover relationships between sets, hovering over any scheme highlights related wireframes (those with intersecting ranges and resulting calculated differences) while hiding others.

\textbf{Justification}.
Throughout our design process for set visualizations, we evaluated two alternatives before selecting wireframes. One approach was to connect the scheme to the layers and use edge bundling; however, this technique introduced significant visual clutter due to the high density of lines.
Alternatively, we experimented with curved edge bands and optimized scheme ordering using a greedy strategy, but despite these optimization efforts, significant overlap persisted, hindering the intuitive perception of set relationships (see the appendix B.3.2).

\subsubsection{Scheme Comparison for Different Model Versions}
Our system supports revision-based editing, where each execution changes the model version (R5).
The preview results in the table (\autoref{fig:teaser}B3) show the editing outcome that would result if an editing scheme were applied to the current model version (indicated in the ``Version'' column).
Initially, the model version is 0. To actually change the underlying model weights (which increments the model version number by 1), users must either right-click on a scheme, or select a layer range (by right-clicking cosine similarity bars or the circles in ranking charts) and then click the ``Edit'' button.
When additional schemes are subsequently added to the table, their ``Version'' column values will be one higher than the previous version (\autoref{fig:teaser}B3).
Users can thus preview the results of different layer selection schemes under the same model version or across different versions.
Clicking the ``Revert'' button returns the model to the previous version (R5).

\subsubsection{Presentation and Comparison of Detailed Results}
Users can expand the detailed editing results for each scheme, which displays the results for all the test prompts and the answers output by the edited model (R4). The results are arranged separately according to the prompts' categories. For the first three categories of prompts, a color-coded background (green for passed and red for failed evaluations) indicates the automatic assessment results, while hovering over a result with the mouse will present the complete results with the model-generated portions highlighted in bold. For ``generation'' type prompts, the system provides a display button for each prompt. Clicking this button triggers the output comparison view (\autoref{fig:teaser}C), which offers a comparative text visualization of the model's output before and after editing for that prompt (R4). This collapsible design enables seamless switching between overview and details (R4).

The output comparison view displays pre-edit and post-edit outputs for a prompt, including one pre-edit output and at least one post-edit output (according to the previously set quantity).
This feature leverages the capability of LLMs to generate multiple different responses of varying quality for a single prompt.
For more detailed analysis, users can select any pair of texts to see a diff visualization that uses a traditional red-green color scheme (red for deletions, green for additions), allowing users to quickly identify key differences between these texts.

\subsection{Checking the Global Impact of Editing}
To help interpret the impact of edits on the global behavior of the model, we reviewed the literature and decided to use the approach of capturing the drift of the hidden states of the last edited layer before and after editing~\cite{fang2025alphaedit}. We input a large number of prompts (usually 1000) into the model and leverage the t-SNE~\cite{maaten2008visualizing} algorithm to project the hidden states corresponding to each prompt onto a two-dimensional plane.

\textbf{Visual Design}. The scatter plot (\autoref{fig:teaser}D) shows the hidden state drift of all prompts before and after editing (R6). The table on the right shows the drift distance of all prompts and the specific output changes before and after editing. These output changes are visualized in the same way as the text visualization in the output comparison view.

\textbf{Interaction}. The scatter plot can be zoomed in and out at any location, allowing users to easily examine points of interest among the large number of data points. Hovering over points displays a label with detailed information, while clicking jumps to the corresponding table row. The data in the right table can be sorted based on the drift values.

\section{Evaluation}\label{sec:5}

\subsection{Usage Scenarios}
We demonstrate the practical value of KEditVis and the insights gained by users through two usage scenarios. The first scenario involves a counterfactual knowledge editing experiment. We chose this scenario considering that COUNTERFACT~\cite{meng2022locating} is a major standardized dataset for knowledge editing.
The second scenario demonstrates knowledge editing as a means of information insertion to correct knowledge errors in the model, representing one of its primary applications.

\begin{figure*}[t]
    \centering
    \includegraphics[width=\textwidth]{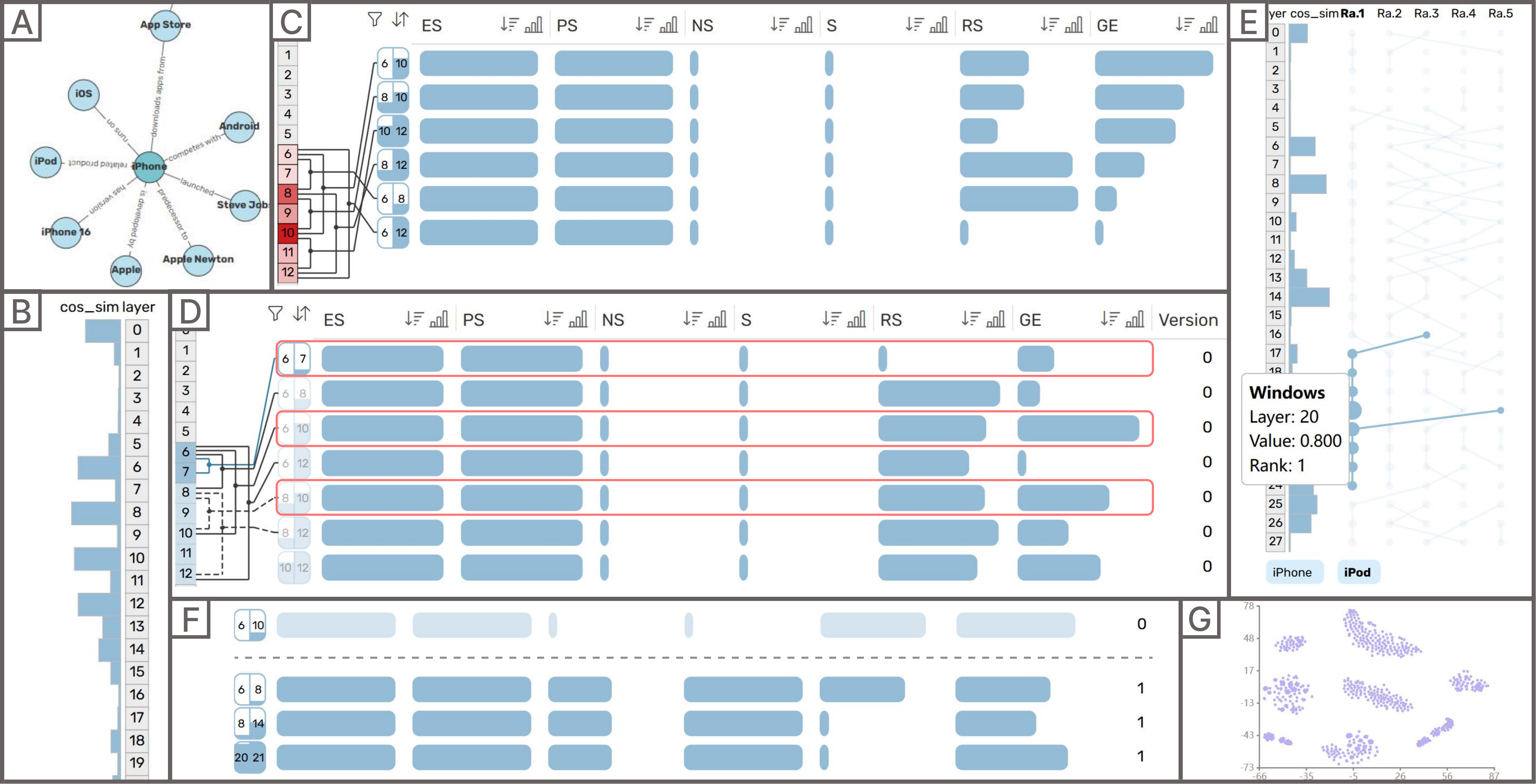}
    \caption{Scenario I: (A) The knowledge graph generated by the keyword ``iPhone''; (B) the cosine similarity bar chart before editing; (C) the descending comparison results after selecting scheme 6-12 and pressing the ``Recommend'' button and the ``Compare'' button; (D) the results after adding scheme 6-7 to the comparison; (E) the cosine similarity chart and ranking chart after editing scheme 6-10; (F) the results after adding schemes 20-21, 8-14, and 6-8 on top of the previously edited weights for scheme 6-10; (G) the scatter plot for checking the global impact of editing.}
    \label{fig:scenario1}
\end{figure*}

\subsubsection{Scenario I: A Counterfactual Experiment}
Jack, an LLM researcher, used KEditVis to uncover insights and inspiration for knowledge editing. Initially, he engaged in conversational interactions with the LLM to be edited. While discussing iPhone related topics, he conceived a counterfactual editing experiment. Through the knowledge graph (\autoref{fig:scenario1}A), he browsed information about iPhone, including its manufacturer, operating system, and related products. Then he introduced a counterfactual statement -- ``iPhone is developed by Microsoft.'' He created a test prompt of efficacy type and had the system generate other types of test prompts automatically. Referring to knowledge graph relationships, he made minor modifications to a few prompts, ultimately obtaining satisfactory test prompts.

Upon double-clicking the fact for editing, the edit view displayed the token ranking chart and the cosine similarities bar chart. Jack observed a significant trend in cosine similarity for layers 6-12 (\autoref{fig:scenario1}B), with prominent bars at layers 6 and 12, thus he added  scheme 6-12 and clicked ``Recommend''. The system immediately suggested reasonable layer-selection schemes within this range.
After clicking ``Compare'', the system visualized knowledge editing results across schemes.
Sorting by descending GE value (\autoref{fig:scenario1}C), he found the bottom-most scheme (6-12), despite being the widest layer range, produced the poorest results. Examining the details of the editing results for scheme 6-12, he discovered the output for a test prompt contained numerous repetitions of the word ``Microsoft.''
Additionally, in the output comparison view, he observed the output for the prompt ``iPhone is a'' displayed repetitive and incoherent sentences.
These characteristics both reflected and corroborated the low GE value.
He concluded that the poor performance was due to over-editing, which demonstrated that simple factual edits may not require editing too many layers to avoid over-editing.

After clicking distribution icons, Jack observed that layers 6-7 exhibited low values in the RS distribution and the lowest values in the GE distribution. He therefore added the scheme 6-7 and the results showed that editing layers 6-7 yielded poor performance (\autoref{fig:scenario1}D). Upon opening the detailed results and the output comparison view, he observed multiple instances of erroneous expressions like ``windows phone'' in the model's outputs, suggesting the model was confusing its inherent knowledge with the edited knowledge, failing to properly integrate the edited information.
Furthermore, he noticed the difference between layers 6-10 and 8-10 consisted precisely of layers 6-7, and observed that the editing results for these two schemes were very close (\autoref{fig:scenario1}D), which led him to conclude that layers 6-7 might neither be critical layers nor suitable layers for editing the current knowledge.

Jack then shifted his attention to the scheme 6-10, which displayed relatively excellent editing results, and right-clicked on it to change model weights to version 1 (model after one edit).
Due to poor NS scores, he examined the test results of neighborhood prompts, and observed the edited model incorrectly attributed the production of iPod to ``Microsoft.'' He then created the fact ``iPod is a product released by Apple'', and upon examining its corresponding ranking chart, he immediately noticing particularly large and prominent circles for layers 20-21 (\autoref{fig:scenario1}E), which belonged to the path for the word ``windows'' (there was strong association between ``windows'' and ``Microsoft''), and the notable cosine similarity trend displayed in layers 8-14 and 6-8 (\autoref{fig:scenario1}E). Based on these observations, he added schemes 20-21, 8-14, and 6-8. After clicking ``Compare'', the results showed scheme 6-8 significantly improved the NS score without noticeably affecting other scores (\autoref{fig:scenario1}F). Now the model correctly identified Apple as the manufacturer of iPod. He recalled that prior work had not investigated re-editing to enhance effects. This experience suggested it might be possible to strengthen edits for specific knowledge by applying a second edit with related information.

Finally, he right-clicked on scheme 6-8 to update to model version 2 and generated a scatter plot. Observing no significant shifts (\autoref{fig:scenario1}G), he zoomed in on points with noticeable changes and used the right-side table to verify no errors in the corresponding outputs.
A few samples related to the editing subject (e.g., electronic products and internet companies) were slightly affected (though the output content was not necessarily incorrect), with no significant impact on the model's global performance, confirming the edits were non-toxic.

\begin{figure}[t]
    \centering
    \includegraphics[width=\columnwidth]{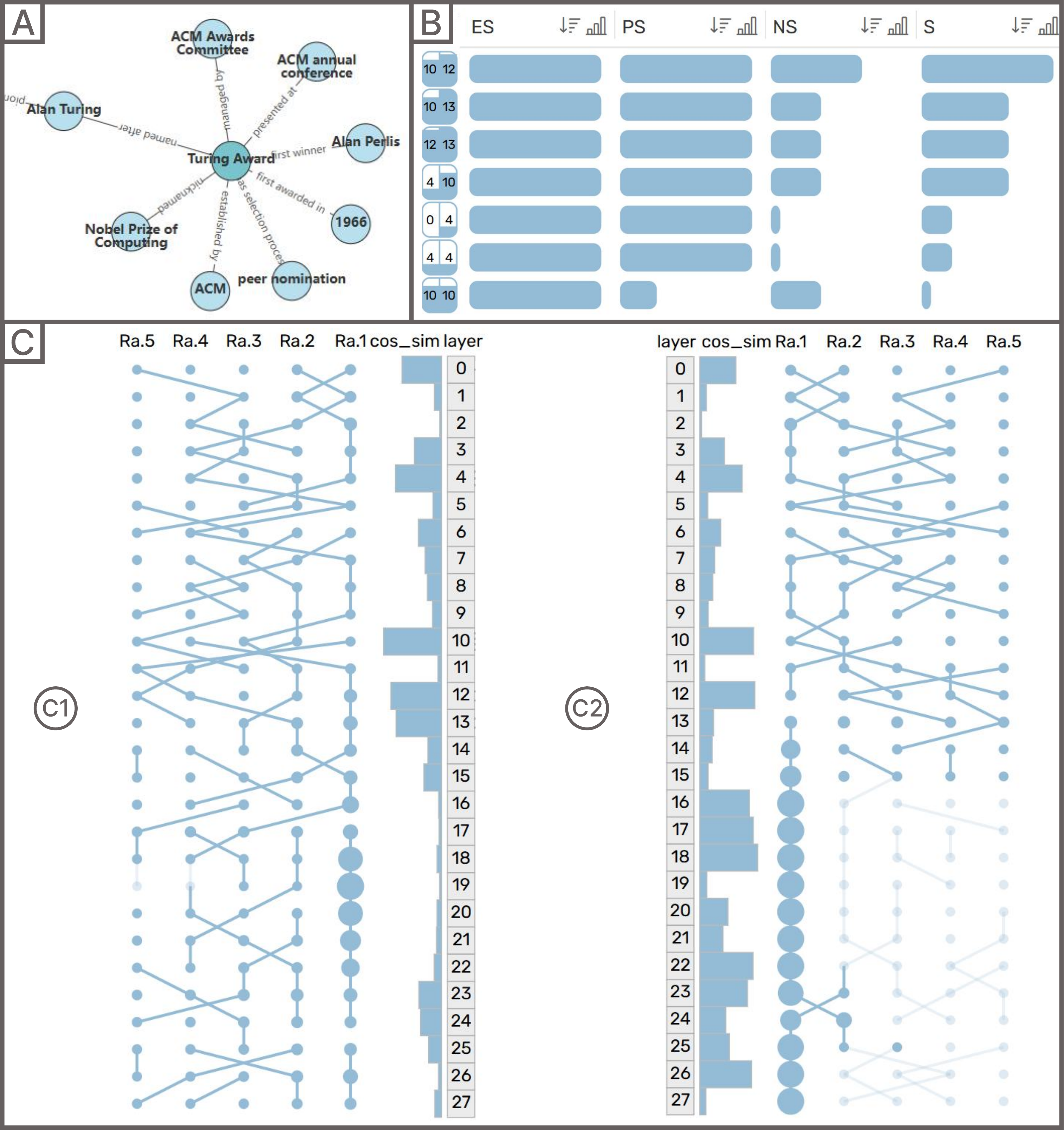}
    \caption{Scenario II: (A) The knowledge graph generated by the keyword ``Turing Award''; (B) the descending sorting of editing results from different layer selection schemes; (C) the ranking chart (for the last token) and the cosine similarity bars, before and after editing scheme 10-12.}
    \label{fig:scenario2}
\end{figure}

\subsubsection{Scenario II: Model Error Correction}\label{sec:5.1.2}
Emily, an LLM expert, applied knowledge editing to correct knowledge in LLMs.
She first analyzed the ``\textit{Turing Award}'' knowledge graph (\autoref{fig:scenario2}A) and inquired about its first recipient, finding the model appeared uncertain about this question, responding with ``\textit{a mystery ...}''. This prompted her to implement editing to instill this specific information into the model.
She generated test prompts using the ``Rewrite'' mode and selectively added them to the list.
Upon discovering from the knowledge graph that the Turing Award is organized by the ACM, she incorporated this relationship as a neighborhood prompt. With minor adjustments, she finalized a list of suitable test prompts.

By examining the edit view (\autoref{fig:scenario2}C1), she noticed a distinct trend in cosine similarity in layers 4-10 and added this scheme.
Next, noticing that layers 10-13 showed both a distinct cosine similarity pattern and high probability values for the key token in the ranking chart, she suspected these layers processed the question's answer and added this scheme. After clicking  ``Recommend'' for more suggested layer selections, she observed a trend for layers 0-4 and added this scheme as well. Finally, out of curiosity, she added single-layer schemes 4 and 10 separately due to their particularly prominent cosine similarity bars.

When viewing the editing results, she sorted by metric S (\autoref{fig:scenario2}B) and found scheme 10-12 showed the most effective editing performance, with layers 10 and 12 having the longest cosine similarity bars. Conversely, individual edits to layers 4 and 10 demonstrated the lowest effectiveness.
She concluded that despite their long cosine similarity bars, single-layer edits produced extremely limited effects, suggesting knowledge representation requires processing through multiple key layers.
She then observed that scheme 0-4 showed poor editing outcomes despite the trend in cosine similarity, while editing the middle layers generally produced more effective results. She concluded that lower layers lacked fully formed specific knowledge representations, so editing them likely interfered with similar knowledge memories of the model, explaining the low NS value for scheme 0-4. Overall, she concluded that single-layer edits were too narrow to capture complete knowledge representation, while broader scopes like scheme 4-10 risked unnecessary interference. Scheme 10-12 offered the excellent balance between precision and coverage.

After choosing scheme 10-12 for editing, which demonstrated optimal editing effectiveness, she examined and compared the ranking charts (for the last token) and cosine similarities before (\autoref{fig:scenario2}C1) and after (\autoref{fig:scenario2}C2) editing.
She divided the model layers into three sections for segmented analysis. (1) In layers 0-9, both charts remained identical, confirming that editing preserved weights in these early layers, resulting in identical hidden states. (2) In layers 10-12, the pre-edit ``honored'' token showed a continuous path, while post-edit it appeared as an isolated point, demonstrating the editing effect; despite this change, for layers 10-12, both the cosine similarity bars and the tokens remained similar between charts, with only token probability distributions shifting, while the target word ``Alan'' only appeared in the final few layers. This pattern verifies how knowledge editing preserves the original information while making subtle modifications to the target layers that gradually accumulate through information propagation in subsequent layers, effectively guiding the final prediction.
(3) In layers 13-27, both ``Turing'' and ``Alan'' maintained high probabilities post-edit, while the overall number of non-zero probability tokens decreased. The final layers showed the emergence of ``Alan'' as the correct answer and exhibited reduced path crossings (transparent paths are ignored as they represent zero probability). All these features demonstrate the model's increased confidence in its prediction after editing.

Finally, she generated a scatter plot and found the point positions had barely changed. After zooming in to carefully examine individual points and the table on the right, she did not discover any issues, confirming the edits' success and leaving her satisfied with the results.

\subsection{Expert Interviews}

We interviewed three knowledge editing experts: E3 and two external experts (E5 and E6).
Our procedure comprised three phases: first, introducing our system through two usage scenario demonstrations; second, allowing experts hands-on system exploration; and finally, conducting interviews for effectiveness and usability feedback, summarized below.

\textbf{Effectiveness}. The experts unanimously found our system useful with effective visualizations. First, they praised the edit view for ``\textit{providing important internal model information previously not easily visible}'' (E5), with its well-designed elements effectively aligned with the data and revealing previously difficult-to-identify patterns.
They also evaluated the edit view holistically, appreciating that it combines result-oriented visualization with internal model representation analysis, with these aspects complementing each other.
Furthermore, the experts agreed that KEditVis's graphics were more intuitive than traditional tabular formats, with set visualization effectively revealing issues like layer range relationships, as E6 noted, ``\textit{I wouldn't have thought of using set visualization to compare different layer choices without the system.}''
Lastly, E6 noted that insights from our scenarios, such as over-editing when modifying too many layers, aligned with findings in his previous research, thus
further validating these insights.

\textbf{Usability}. The experts praised KEditVis's usability, confirmed its utility in helping them conduct their work and revealing previously unnoticed insights, and expressed interest in adoption across research and engineering contexts.
They would use KEditVis to ``\textit{explore editing schemes and evaluate edits in research, while using the knowledge graph to detect errors and verify edits in engineering contexts}'' (E5).

\textbf{Expert Reflections}.
First, the experts noted re-editing the iPod manufacturer successfully improved NS and S metrics despite slight decreases in RS and GE. They still acknowledged its effectiveness, since different applications may prioritize various metrics differently.
Second, they speculated ranking chart patterns -- reduced path crossings in the latter layers and prominently singular high-probability tokens -- likely occur commonly when editing addresses LLMs' lack of clear, definitive information about specific topics.

\subsection{User Study}
We conducted a task-based user study to evaluate our system’s usability with 12 participants (P1-P12), all proficient with LLMs.
Six were knowledge editing experts, three had basic understanding, and three had no prior knowledge editing experience.
None of them were involved in the development of KEditVis or the expert interviews.

The user study consisted of three stages. The first stage was an introduction, where the moderator presented background knowledge and use cases of KEditVis to participants. The second stage required participants to complete three different knowledge editing tasks, each following the system's workflow. The first and third tasks were derived from the two usage scenarios described earlier, involving the editing of iPhone manufacturers and Turing Award recipients, respectively, while the second task focused on editing information related to U.S. presidents. During this process, we encouraged users to think aloud~\cite{fonteyn1993description} about their analytical reasoning and select what they considered the optimal editing scheme at the end. In the third stage, we asked participants to complete a System Usability Scale (SUS)~\cite{brooke1996SUS} questionnaire using a 5-point Likert scale and conducted interviews with them.
For detailed task descriptions and questionnaires, please refer to the appendix D.

\subsubsection{Quantitative Results}
\begin{figure}[!ht]
    \centering
    \includegraphics[width=\columnwidth]{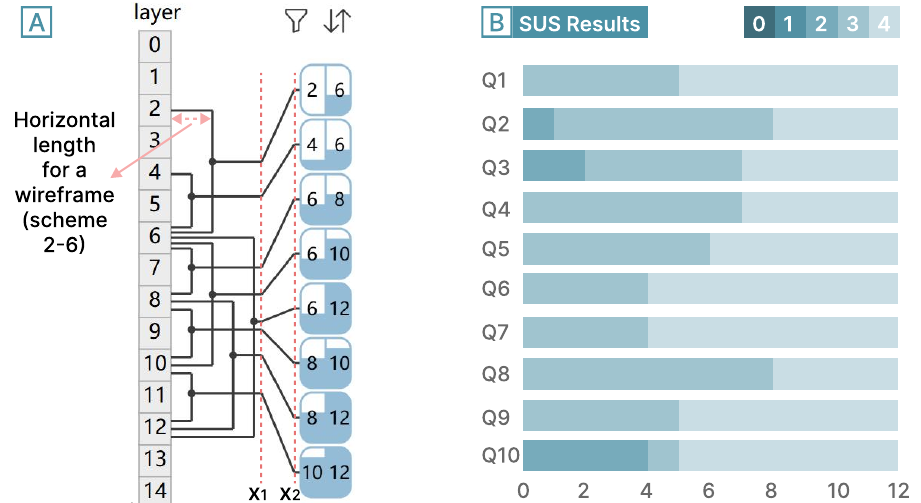}
    \caption{(A) Diagram of the set visualization we designed; (B) results of the SUS questionnaire in the user study.}
    \label{Quantitative-results}
\end{figure}

All participants completed all tasks successfully. Each participant tried an average of 5.8 schemes per task. Participants gave a SUS score of 86.25, which exceeds the 80.3 threshold for the top 10\%~\cite{brooke1996SUS}.
Based on SUS factor research~\cite{lewis2009Factor}, we calculated usability and learnability scores of 86.98 (Q1-3, Q5-9) and 83.33 (Q4, Q10), respectively.
As shown in the detailed results in \autoref{Quantitative-results}B, participants generally evaluated the system's usability and learnability positively.
However, a few participants found certain system features too implicit, noting that there was ``\textit{no explicit button for expanding and collapsing detailed results}'' (P5), and that ``\textit{some background knowledge needed to be learned, such as the principles of layer selection methods}'' (P1, P5, P10, P12).

\textbf{Post-hoc Analysis.}
We recorded the best editing scheme selected by each participant for each task and compared their editing effects against three baselines, using the same testing benchmark. Baseline I used fixed layers from current mainstream knowledge editing methods~\cite{meng2022mass}, while Baseline II used automated approaches to choose a layer selection scheme for editing~\cite{huang-etal-2024-commonsense}. Baseline III first applied Baseline II's methodology to identify the six most suitable layer selection schemes, then selected the optimal scheme from these candidates based on editing performance.
The reason for selecting these baselines is to demonstrate that our interactive method is effective for editing and provides value compared to both fixed-layer and automated layer selection approaches.

The comparison was conducted with Student's t-test.
The results below revealed that the schemes selected by participants significantly outperformed all baselines, with only a few instances showing the same or comparable performance, thus confirming the effectiveness of human participation in layer selections. While automated layer selection methods do not achieve optimal results in all cases, human decision-making can effectively compensate for their deficiencies.

\textbf{Task 1}. Participant-selected schemes ($M = 0.683$, $SD = 0.023$) significantly outperformed Baseline I ($t(11) = 44.000$, $p < 0.001$, mean difference $MD = 0.293$, $95\%\ CI\ [0.279, 0.308]$, $d = 12.702$) and Baseline II ($t(11) = 11.000$, $p < 0.001$, $MD = 0.073$, $95\%\ CI\ [0.059, 0.088]$, $d = 3.175$), while performing comparably to Baseline III.

\textbf{Task 2}. Participant-selected schemes ($M = 0.747$, $SD = 0.012$) significantly outperformed Baseline I ($t(11) = 23.000$, $p < 0.001$, $MD = 0.077$, $95\%\ CI\ [0.069, 0.084]$, $d = 6.640$), Baseline II ($t(11) = 11.000$, $p < 0.001$, $MD = 0.037$, $95\%\ CI\ [0.029, 0.044]$, $d = 3.175$) and Baseline III ($t(11) = 11.000$, $p < 0.001$, $MD = 0.037$, $95\%\ CI\ [0.029, 0.044]$, $d = 3.175$).

\textbf{Task 3}. Participant-selected schemes ($M = 0.692$, $SD = 0.034$ in RS; $M = 5.243$, $SD = 0.307$ in GE) performed comparably to Baseline I but significantly outperformed both Baseline II ($t(11) = 7.231$, $p < 0.001$, $MD = 0.072$, $95\%\ CI\ [0.050, 0.093]$, $d = 2.087$) and Baseline III ($t(11) = 7.231$, $p < 0.001$, $MD = 0.072$, $95\%\ CI\ [0.050, 0.093]$, $d = 2.087$) on metric RS. They significantly outperformed Baseline I ($t(11) = 6.813$, $p < 0.001$, $MD = 0.603$, $95\%\ CI\ [0.408, 0.798]$, $d = 1.967$), Baseline II ($t(11) = 5.232$, $p < 0.001$, $MD = 0.463$, $95\%\ CI\ [0.268, 0.658]$, $d = 1.510$) and Baseline III ($t(11) = 5.232$, $p < 0.001$, $MD = 0.463$, $95\%\ CI\ [0.268, 0.658]$, $d = 1.510$) on metric GE, as well as performing comparably to all baselines on metric S.

Furthermore, we conducted a cosine similarity analysis between pre-edit and post-edit hidden states, computed before t-SNE projection.
The results showed that all cosine similarities for models edited by both baselines and participants are very close to 1 (>0.99), confirming that the edits have negligible global impact on model performance (see the appendix D.3).
We also evaluated whether model outputs showed semantic changes before and after editing by examining 1000 samples in the drift view.
The results showed that the outputs for most samples did not change, with only a small number showing changes that mostly preserved semantic meaning (see the appendix D.4).

\subsubsection{Qualitative Results}\label{sec:5.3.2}
We analyzed participants' utterances from think-aloud and interview sessions, as summarized below, to evaluate KEditVis's effectiveness.

\textbf{KEditVis effectively helps weigh and select appropriate layers for editing (R2).}
Cosine similarity charts and ranking charts helped participants make trade-offs and increased confidence in their choices.
Participants reported selecting layers where both charts showed overlapping trends, and excluded layers with short cosine similarity bars despite high target token probability, or those with low target token probability despite long cosine similarity bars.
They also selected layers showing clear cosine similarity trends when target tokens (in the ranking chart for the last token) had not yet reached high probability values, as these layers represent a critical stage where knowledge representations are forming rather than fully established.
Editing at this stage is often more effective than editing after knowledge forms.

\textbf{KEditVis effectively helps analyze changes in the model before and after editing (R2).}
Participants compared ground truth token probabilities before and after editing, and probability changes between old and new answers.
When comparing pre- and post-editing ranking charts, they observed that successful edits maintained consistency in where the target token initially formed and its general trajectory. This pattern could potentially indicate the toxicity of edits.
Through comparative visualizations, they also found minimal cosine similarity changes in edited layers but significant changes in subsequent layers, which P2 and P7 attributed to greater involvement of related tokens.

\textbf{KEditVis effectively helps compare and analyze different schemes (R3).} Participants reported they could confidently find excellent editing schemes.
When unable to distinguish among a few editing results, they favored schemes with fewer layers while comparing cosine similarities and token probability changes to support the final decisions.
They discovered that PS values were strongly influenced by the number of key layers selected, and modifying middle layers readily impacted NS scores, which P5 attributed to middle layers' coupling and converging information. Conversely, modifying later layers affected GE values, as these layers organize language and prepare the final output.

\textbf{KEditVis effectively presents editing results at both overview and detail levels (R4).}
Participants found the system's combination of overall metrics and detailed displays useful since automated assessments can be inaccurate.
They examined detailed results when noticing metric anomalies, and consulted detailed views when comparing schemes with similar metrics for finer analysis.
Text visualization in the output comparison helps identify primary modifications. P2 discovered that editing different layers created distinct knowledge associations, suggesting layer selections can consider where target tokens and replacement tokens exist for a more comprehensive approach.

Additionally, all participants found the system-generated test prompts satisfactory (R1), with only two participants (P2, P11) making minor changes. Several participants (P6, P9, P12) spontaneously proposed re-editing the best editing scheme in the third task to improve a few metrics (R5), aligning with our scenario I. Participants noted the drift view was practical, flexible, and effective for assessing global edit impacts (R6), and all indicated that no negative impact of the edits on the model's global performance was found.
They also indicated that due to the efficiency of knowledge editing itself, KEditVis's computational cost is far lower than traditional fine-tuning, with timely system response (refer to the appendix E for details about runtime cost). Although it requires certain human efforts for decision making, it effectively helps analyze editing schemes and obtain useful insights.

\section{Discussion}

\textbf{Implications}.
First, our layer-based approach reveals hidden patterns that advance knowledge editing research. Unlike traditional approaches focusing only on algorithms and overall metrics, our interactive method enables users to easily understand editing impacts, identify issues, and design effective strategies through sequence analysis, comparative analysis, and case-by-case examination.
Second, our interactive visualization helps users analyze and select editing layers, identify imperfect results, and perform effective edits. Automated methods have limitations -- the range automatically identified by minimum cosine similarity values is not always optimal, and target tokens sometimes do not appear in ranking charts.
These limitations can be addressed through human-in-the-loop intervention, enabling contextual judgment, introducing domain expertise, and allowing nuanced decision-making beyond automated metrics.
Third, while previous research~\cite{hong2024interpretability, huang-etal-2024-commonsense} demonstrated the effectiveness of cosine similarity and token projection approaches across datasets, our work revealed that relying on single selection methods can be unreliable for certain instances. This finding suggests combining various selection methods for different instances.

\textbf{Generalizability}.
Our visual analytics approach is generalizable and extends to different editing methods and LLMs, with our contributions independent of specific editing techniques or LLMs.
Our system supports different knowledge editing backends~\cite{wang-etal-2024-easyedit}. Local modification methods -- currently mainstream -- may differ in algorithmic details but all follow the locate-then-edit strategy and remain compatible with our system.
Our system is applicable to classical Transformer-based LLMs, such as the GPT family.
Knowledge editing methods inherently support different LLMs because they operate on the fundamental architectural components shared by most LLMs, without depending on specific implementations~\cite{fang2025alphaedit, meng2022mass}.
Layer selection methods also utilize the commonalities between different LLMs, supporting various LLM architectures~\cite{hong2024interpretability, huang-etal-2024-commonsense}.
In addition, our approach can extend beyond knowledge editing, such as model distillation. Through visual comparison, we can determine which layers contain critical knowledge that should be prioritized for retention in distilled models.

\textbf{Limitations and Future Work}.
First, due to human involvement, it is difficult to measure whether KEditVis and the baselines invest completely equivalent efforts in our user study, as there is currently no approach to measure whether humans and machines invest equal efforts. More refined quantitative comparisons can be conducted to further enhance rigor in the future.
Second, as the number of edited facts and model layers increases, more schemes may need to be attempted, leading to higher computational costs and user efforts. Future work could explore efficient scheme recommendation methods.
Third, future work could compare different editing methods across LLMs to identify optimal approaches and explore neuron-level editing for finer control.

\section{Conclusion}

This paper presents KEditVis, a novel visual analytics system for LLM knowledge editing that helps users target appropriate model layers, identify reasons for suboptimal edits, and improve outcomes effectively.
Comprising usage scenarios, expert interviews, and a user study, our evaluation demonstrates its effectiveness and reveals valuable insights.
As knowledge editing has become an active research area, this work addresses current workflow limitations through an interactive approach, contributing to the advancement of knowledge editing methodologies.

\acknowledgments{
The authors are grateful to the reviewers for their thorough reviews that helped strengthen this work.
This work was supported by Zhejiang Provincial Natural Science Foundation of China under Grant No. LD25F020003,  and NSFC (62402421, U22A2032, 62421003).
}

\newpage

\bibliographystyle{abbrv-doi-hyperref}
\bibliography{check}

\end{document}